\documentstyle[12pt,epsf]{article}
\def\abstracts#1{{
	\centering{\begin{minipage}{30pc}\tenrm\baselineskip=12pt\noindent
	\centerline{\tenrm ABSTRACT}\vspace{0.3cm}
	\parindent=0pt #1
	\end{minipage}}\par}}
 1
 1
 1

\font\tenbf=cmbx10
\font\tenrm=cmr10
\font\tenit=cmti10

\def\ra{\rightarrow}
\def\prd#1#2#3{{\it Phys. Rev.} {\bf D#1} #2 (19#3)}
\def\pl#1#2#3{{\it Phys. Lett.} {\bf #1B} #2 (19#3)}
\def\np#1#2#3{{\it Nucl. Phys.} {\bf B#1} #2 (19#3)}
\def\prl#1#2#3{{\it Phys. Rev. Lett.} {\bf #1} #2 (19#3)}
\def\beq{\begin{equation}}
\def\eeq{\end{equation}}

\def\beqn{\begin{eqnarray}}
\def\eeqn{\end{eqnarray}}
\begin{document}
\def\ba{\begin{array}}
\def\ea{\end{array}}

\centerline{\tenbf PROBING ANOMALOUS GAUGE BOSON }
\baselineskip=16pt
\centerline{\tenbf COUPLINGS  AT LEP\footnote{
Invited talk given by S.D. at the {\it Joint U.S.-Polish Workshop on
Physics from Planck Scale to Electroweak Scale}, Sept 21-24, 1994,
Warsaw, Poland and by G.V. at the {\it 1er. Taller
Colombiano de Fenomenologia Meeting}, Aug. 16-18,1994,
Albuquerque, New Mexico.  }}
\vspace{0.8cm}
\centerline{\tenrm S.~DAWSON}
\baselineskip=13pt
\centerline{\tenit Physics Department, Brookhaven National Laboratory,
Upton, N.~Y.~~11973}
\vspace{0.3cm}
\centerline{\tenrm and}
\vspace{0.3cm}
\centerline{\tenrm G.~VALENCIA}
\baselineskip=13pt
\centerline{\tenit Department of Physics, Iowa State University,
Ames, IA~~50011}
\vspace{0.9cm}
\abstracts{
We bound anomalous gauge boson couplings using LEP data
for the $Z\rightarrow {\overline f} f$ partial widths.
We use an effective field theory formalism
to compute the one-loop corrections
resulting from non-standard model three and four gauge boson
vertices. We find that   measurements at LEP constrain the three
gauge boson couplings at a level comparable to that
 obtainable at LEPII.}
\vfil
\rm\baselineskip=14pt
\section{Introduction}

High precision measurements at the $Z$ pole at LEP
have been used to place stringent limits on new physics beyond the standard
model.
Under the assumption that the dominant effects of the new physics
would show up as corrections to the gauge boson self-energies, the
LEP measurements have been used to parameterize the possible new
physics in terms of three observables, $S$, $T$, $U$.\cite{pestak,alta}
A fourth observable
corresponds to the partial width $Z \ra b \overline{b}$.

In view of the extraordinary agreement between the standard model
predictions and the observations, it seems reasonable to assume
that the $SU(2)_L \times U(1)_Y$ gauge theory of electroweak interactions
is essentially correct and that the only sector of the theory
lacking experimental support is the symmetry breaking sector. There
are many extensions of the minimal standard model that incorporate
different symmetry breaking possibilities. One large class of models
is that in which the interactions responsible for the symmetry breaking
are strongly coupled and their
effects would show up in experiments as deviations from the minimal
standard model couplings.
In this paper we use the measurements of the  partial decay widths
of the $Z$ boson to place bounds on anomalous gauge boson couplings.

\section{Formalism}

We assume that the electroweak interactions are given
by an $SU(2)_L \times U(1)_Y$ gauge
theory with spontaneous symmetry breaking to $U(1)_{EM}$ and
that we do not have any information on the symmetry breaking sector
except that it is strongly interacting and that any new particles have
masses higher than several hundred GeV. This
scenario can be described with an effective Lagrangian with operators
organized according to the number of derivatives or gauge fields
they have. If we call $\Lambda$ the scale at which the symmetry breaking
physics comes in, this organization of operators corresponds to
an expansion of amplitudes in powers of $(E^2~{\rm or}~v^2)/\Lambda^2$.
The lowest order effective Lagrangian for the symmetry breaking
sector of the theory is: \cite{longo}
\beq
{\cal L}^{(2)}={v^2 \over 4}{\rm Tr}\biggl[D^\mu \Sigma^\dagger D_\mu
\Sigma \biggr].
\label{lagt}
\eeq
In our notation $W_{\mu}$ and $B_{\mu}$ are
the $SU(2)_L$ and $U(1)_Y$  gauge fields with
$W_\mu \equiv W^i_\mu \tau_i$.
The matrix $\Sigma \equiv \exp(i\vec{\omega}\cdot \vec{\tau} /v)$, contains the
Goldstone bosons $\omega_i$ that give the $W$ and $Z$ their
masses via the Higgs mechanism and the $SU(2)_L \times U(1)_Y$
covariant derivative is given by:
\beq
D_\mu \Sigma = \partial_\mu \Sigma +{i \over 2}g W_\mu^i \tau^i\Sigma
-{i \over 2}g^\prime B_\mu \Sigma \tau_3.
\label{covd}
\eeq
This non-renormalizable Lagrangian
is interpreted as an effective field theory, valid below
some scale $\Lambda \leq 3$~TeV. The lowest order interactions between
the gauge bosons and fermions, as well as the kinetic energy
terms for all fields,  are the same as those in the minimal
standard model.

For LEP observables, the operators that can appear at tree-level
are those that modify the gauge boson self-energies. To order
${\cal O}(1/\Lambda^2)$ there are only three:\cite{longo,appel}
\beq
{\cal L}^{(2GB)}=\beta_1{v^2\over 4} \biggl({\rm Tr} \biggl[
\tau_3 \Sigma^\dagger D_\mu \Sigma\biggr]\biggr)^2
+ \alpha_8g^2 \biggl({\rm Tr}\biggl[\Sigma \tau_3 \Sigma^\dagger
W_{\mu \nu}\biggr]\biggr)^2
 + g g^{\prime}{v^2 \over \Lambda^2} L_{10}\, {\rm Tr}\biggl[ \Sigma
B^{\mu \nu}
\Sigma^\dagger W_{\mu \nu}\biggr],
\label{oblique}
\eeq
which contribute respectively to $T$, $U$ and $S$.

In this paper we will consider operators that affect the $Z$ partial
widths at the one-loop level. We will restrict our study to only those
operators that appear at order ${\cal O}(1/\Lambda^2)$ in the gauge-boson
sector and that respect the custodial symmetry in the limit
$g^\prime \ra 0$.  We will also consider an additional operator
which is parity violating, but CP conserving.   Because of the parity
violating nature of this term it can lead to observable signals
at LEPII via $e^+e^-\rightarrow W^+W^-$.\cite{gfive}
The Lagrangian we consider is thus:
\beqn
{\cal L}^{(4)}\ &=&\ {v^2 \over \Lambda^2}  \biggl\{ L_1 \,
\biggl( {\rm Tr}\biggl[
D^\mu\Sigma^\dagger D_\mu \Sigma \biggr]\biggl)^2
\ +\  L_2 \,\biggl({\rm Tr}\biggl[
 D_\mu\Sigma^\dagger D_\nu \Sigma\biggr]\biggl)^2
 \nonumber \\
&& - i g L_{9L} \,{\rm Tr}\biggl[W^{\mu\nu} D_\mu
\Sigma D_\nu \Sigma^\dagger\biggr]
\ -\ i g^{\prime} L_{9R} \,
{\rm Tr}\biggl[ B^{\mu \nu}
D_\mu \Sigma^\dagger D_\nu\Sigma\biggr]\nonumber \\
&&+ g {\hat \alpha} \epsilon^{\alpha \beta\mu\nu}
Tr\biggl(\tau_3 \Sigma^\dagger D_\mu \Sigma\biggr)
Tr\biggl(W_{\alpha \beta} D_\nu \Sigma\Sigma^\dagger\biggr)
\biggr\},
\label{lfour}
\eeqn
where the field strength tensors are given by:
\beqn
W_{\mu\nu}&=&{1 \over 2}\biggl(\partial_\mu W_\nu -
\partial_\nu W_\mu + {i \over 2}g[W_\mu, W_\nu]\biggr)
\nonumber \\
B_{\mu\nu}&=&{1\over 2}\biggl(\partial_\mu B_\nu-\partial_\nu B_\mu\biggr)
\tau_3 .
\label{fsten}
\eeqn

The operators in Eq.~\ref{oblique} and Eq.~\ref{lfour} would
arise when considering the effects
of those in Eq.~\ref{lagt} at the one-loop level or
from the new physics responsible for symmetry
breaking at a scale $\Lambda$ at order $1/\Lambda^2$. We therefore
explicitly introduce the factor $v^2/\Lambda^2$ in our definition
of ${\cal L}^{(4)}$ so that the coefficients
$L_i$ are naturally of ${\cal O}(1)$.
In the present paper we will compute the
contribution  from the operators
of Eqs. \ref{oblique} and \ref{lfour}
to the $Z$ partial widths that are measured at LEP.

We will first perform a complete calculation to order ${\cal O}
(1/\Lambda^2)$. That is, we will include the one-loop contributions from
the operator in Eq.~\ref{lagt} (and gauge boson kinetic energies).
The divergences generated in this calculation are absorbed by
renormalization of the couplings in Eq.~\ref{oblique}. This calculation
will illustrate our method and as an example we use it to place
bounds on $L_{10}$.

We will then place bounds on the couplings of Eq.~\ref{lfour} by considering
their one-loop effects. The divergences generated in this one-loop calculation
would be removed in general by renormalization of the couplings in the
${\cal O}(1/\Lambda^4)$ Lagrangian of those operators that modify the
gauge boson self-energies at tree-level and perhaps by additional
renormalization of the couplings in Eq.~\ref{oblique}.
Interestingly,
we find that we can obtain a completely finite result for the $Z \ra
\overline{f} f$ partial widths using only the operators in Eq.~\ref{oblique}
as counterterms when we consider the parity conserving operators.
This is not the case, however, for the parity violating operator,
${\hat \alpha}$.

We first regularize the
integrals in $n$ space-time dimensions and remove all the poles in
$n-4$ as well as the finite analytic terms by a suitable definition of
the renormalized couplings. We base our
analysis on the leading non-analytic
terms proportional to $L_i \log \mu$. These terms determine the
running of the $1/\Lambda^4$ couplings and cannot be generated by
tree-level terms at that order.
 With a carefully chosen renormalization scale $\mu$
( such that the logarithm is of order one), these terms give us
the correct order of magnitude for the size of the $1/\Lambda^4$
coefficients.\cite{georgi}
Quadratic divergences will be absorbed by the new physics arising at
${\cal O}(1/\Lambda^4)$ and cannot be used to limit the coefficients of the
${\cal O}(1/\Lambda^2)$ Lagrangian.\cite{quaddiv}
Clearly, the LEP observables do not
measure the couplings in Eq.~\ref{lfour} and it is only from
naturalness arguments like the one above that we can place bounds
on the anomalous gauge-boson couplings.
{}From this perspective, it is clear that these bounds are not a substitute
for direct measurements in future high energy machines.

We will perform our calculations in unitary gauge, so we set $\Sigma=1$
in Eqs.~\ref{lagt}, \ref{oblique} and \ref{lfour}.
For the lowest order operators we use the conventional
input parameters: $G_F$ as measured in muon decay;
the physical $Z$ mass: $M_Z$; and $\alpha(M_Z)=1/128.8$. Other lowest
order parameters are derived quantities and we adopt one of the
usual definitions for the mixing angle:
\beq
s_Z^2 c_Z^2 \equiv{\pi \alpha(M_Z)\over \sqrt{2} G_F M_Z^2}.
\label{szdef}
\eeq

We neglect the mass and momentum of the external fermions
compared to the $Z$ mass.
The only fermion mass that is kept in our calculation is
the mass of the top-quark when it appears as an intermediate state.

With this formalism we proceed to compute the $Z\ra f \overline{f}$
partial width from the following ingredients.

\begin{itemize}

\item The $Z\ra f {\overline f}$ vertex, which we write as:
\beq
i {\Gamma}_\mu =  -i{e\over 4 s_Z c_Z}
\gamma_\mu\biggl[
(r_f +\delta r_f)(1+\gamma_5) +(l_f+\delta l_f)
 (1-\gamma_5)\biggr ]
\label{vertex}
\eeq
where $r_f=-2Q_f  s_Z^2$ and $l_f=r_f+T_{3f}$.
The terms $\delta l_f$ and $\delta r_f$ occur at one-loop both at order
$1/\Lambda^2$ and at order $1/\Lambda^4$ and can be found
in Ref. [9].

\item The renormalization of the lowest order input parameters.
At order $1/\Lambda^2$ it is induced by tree-level anomalous
couplings and one-loop diagrams with lowest order vertices. At
order $1/\Lambda^4$ it is induced by one-loop diagrams with an
anomalous coupling in one vertex.

\item Tree-level and one-loop contributions to $\gamma Z$ mixing.

\item Wave function renormalization.

\end{itemize}

With all these ingredients we can find our
 final expression for the physical partial width. We find:
\beq
\Gamma(Z\ra f \overline{f}) \equiv \Gamma_{SM}(Z\ra f\overline{f})
\biggl( 1 + {\delta \Gamma_f^{L_i} \over \Gamma_0(Z\ra f\overline{f})}\biggr).
\label{defw}
\eeq
where $\Gamma_0$ is the lowest order tree level result,
\beq
\Gamma_0 (Z\rightarrow f {\overline f})=
N_{cf} (l_f^2 + r_f^2) {G_F M_Z^3 \over 12 \pi \sqrt{2}}.
\label{widthl}
\eeq
The values for $\delta L_f^{L_i}$ are given in the following sections.
Using Eq.~\ref{defw} we introduce
additional terms proportional to products of standard model one-loop
corrections and corrections due to anomalous couplings. These are
small effects that do not affect our results.

We will not attempt to obtain a global fit to the parameters in our
formalism from all possible observables. Instead we use the
partial $Z$ widths:\cite{lepex}
\beqn
\Gamma_e &=&  83.98\pm 0.18 ~ MeV
\nonumber \\
\Gamma_\nu &=& 499.8 \pm 3.5~ MeV
\nonumber \\
\Gamma_Z &=&2497.4 \pm 3.8~MeV
\nonumber \\
R_h &=& 20.795\pm 0.040~
\label{data}
\eeqn
The bounds on new physics are obtained by subtracting the standard
model predictions at one-loop from the measured partial widths
as in Eq.~\ref{defw}. We
use the numbers of Langacker \cite{smpred} which use the global
best fit values for $M_t$ and $\alpha_s$ with $M_H$ in the range
$60-1000$~GeV. The first error is from the uncertainty in $M_Z$ and
$\Delta r$, the second is from $M_t$ and $M_H$, and the one in
brackets is from the uncertainty in $\alpha_s$.\cite{smpred}
\beqn
\Gamma_e &=& 83.87~\pm 0.02 \pm 0.10~ MeV
\nonumber \\
\Gamma_\nu &=& 501.9 \pm 0.1 \pm 0.9~ MeV
\nonumber \\
\Gamma_Z &=& 2496 \pm 1 \pm 3 \pm [3]~ MeV
\nonumber \\
R_h &=& 20.782 \pm 0.006 \pm 0.004 \pm[0.03]
\label{theory}
\eeqn
We add all errors in quadrature.
\footnote{
See also  the contribution of A. Sopczak\cite{as}   to these
proceedings for a discussion of extracting bounds on
new physics from LEP measurements.  }

\section{Results}

In this section we compute the corrections to the $Z \ra f \overline{f}$
partial widths from the couplings of Eq.~\ref{lfour}, and compare them
to recent values measured at LEP. We treat each coupling constant
independently, and compute only its {\it lowest} order contribution to the
decay widths.

\subsection{Bounds on $L_{10}$ at order $1/\Lambda^2$.}

The operators in Eq.~\ref{oblique} are the only ones that induce a
tree-level correction to the gauge boson self-energies to
order ${\cal O}(1/\Lambda^2)$.
These corrections are  of course well known and correspond,
{\it at leading order}, to the new physics
contributions to $S,~T,~U$.\cite{pestak}

We have regularized our
one-loop integrals in $n$ dimensions and isolated the  ultraviolet
poles $1/\epsilon = 2/(4-n)$. We find that we obtain a finite answer
to order $1/\Lambda^2$ if we adopt the following renormalization
scheme:
\beqn
{v^2 \over \Lambda^2} L^r_{10}(\mu) &=& {v^2 \over \Lambda^2}
L_{10}-{1 \over 16 \pi^2} {1 \over 12}
\biggl( {1 \over \epsilon} + \log{\mu^2 \over M_Z^2} \biggr)
\nonumber \\
\beta_1^r(\mu) &=& \beta_1 -
{e^2 \over 16 \pi^2}{3 \over 2 c^2_Z}
\biggl( {1 \over \epsilon} + \log{\mu^2 \over M_Z^2} \biggr).
\label{renorm}
\eeqn
We thus replace the bare parameters $L_{10}$ and $\beta_1$ with the
scale dependent ones above. As a check of our answer, it is interesting
to note that we would also obtain a finite answer by adding to our
result  the one-loop contributions to the self-energies
obtained in unitary gauge in the minimal standard model with one Higgs
boson in the loop.

Our result for $L_{10}$ at order $1/\Lambda^2$ is then:
\beq
{\delta \Gamma_f^{L_{10}} \over \Gamma_0(Z\ra f\overline{f})}=
{e^2 \over c_Z^2 s_Z^2}L^r_{10}(\mu)
{v^2 \over \Lambda^2}{2r_f(l_f +r_f) \over
l_f^2+r_f^2}{c_Z^2 \over s_Z^2-c_Z^2}
\label{shten}
\eeq
Once again we point out that, at this order, the contribution of
$L_{10}$ to the
LEP observables occurs only through modifications to the self-energies
that are proportional to $q^2$. At this order it is therefore
possible to identify the effect of $L_{10}$ with the oblique
parameter $S$. If we were to compute the effects
of $L_{10}$ at one-loop (as we do for the $L_{9L,9R}$) comparison with
$S$ would not be appropriate.

\begin{figure}[htb]
\centerline{\hfil\epsffile{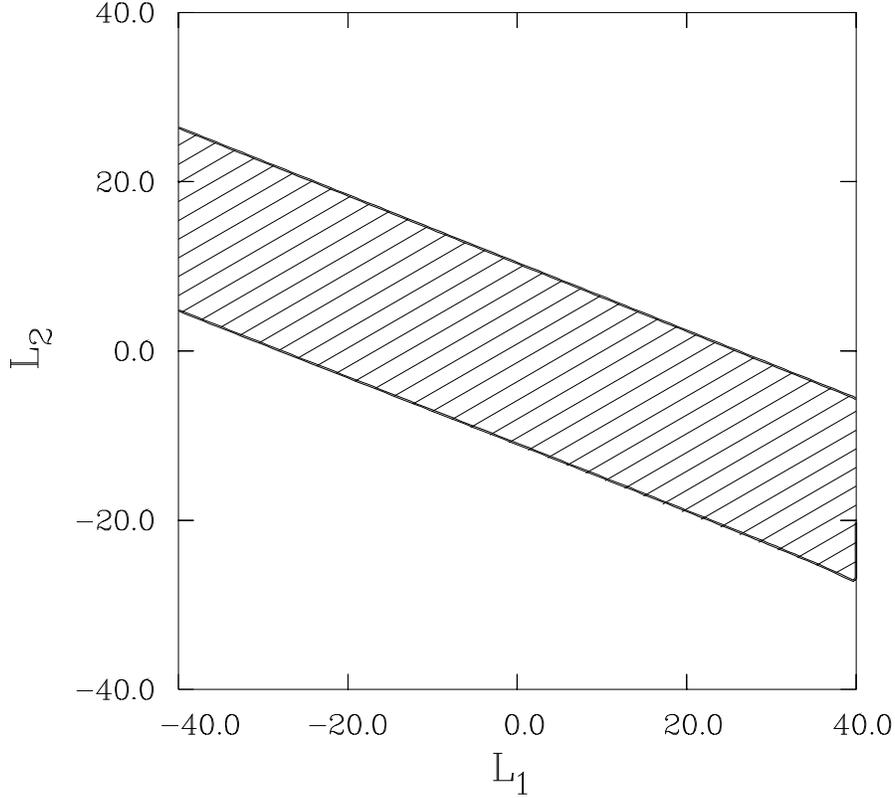}\hfil}
\caption[]{$90\%$ confidence level bounds on $L_{1,2}$  from the $Z\rightarrow
f {\overline f}$ partial widths, (Eq. 14).  The allowed region is shaded.}
\end{figure}

Numerically we find the following $90\%$ confidence level bounds on $L_{10}$
when we take the scale $\Lambda=2$~TeV:
\beqn
\Gamma_e &\ra & -1.7 \leq L^r_{10}(M_Z)_{new} \leq 3.3 \nonumber \\
R_h &\ra & -1.5 \leq L^r_{10}(M_Z)_{new} \leq 2.0 \nonumber \\
\Gamma_Z &\ra & -1.1 \leq L^r_{10}(M_Z)_{new} \leq 1.5
\label{tennum}
\eeqn

\subsection{Bounds on $L_{1,2}$ at order $1/\Lambda^4$.}

The couplings $L_{1,2}$ enter the one-loop
calculation of the $Z \ra f \overline{f}$
width through the  four gauge boson couplings.
Our prescription
calls for using only the leading non-analytic contribution
to the process $Z \ra f \overline{f}$. This contribution can be extracted
from the coefficient of the pole in $n-4$.
Care must be taken to isolate the poles of
ultraviolet origin (which are the only ones that interest us) from those
of infrared origin that appear in intermediate steps of the calculation
but that cancel as usual when one includes real emission processes as well.

Since in unitary gauge $L_{1,2}$ modify only the four-gauge
boson couplings at the one-loop level, they
enter the calculation of the $Z$ partial widths only through the
self-energy corrections and the renormalization of
the lowest order couplings.
These operators induce a non-zero value for $\Delta \rho \equiv
\Pi_{WW}(0) / M_W^2 - \Pi_{ZZ}(0)/ M_Z^2$. For the observables
we are discussing, this is the {\it only} effect of $L_{1,2}$.
We do not place bounds on them from global fits of the oblique
parameter $T$, because we have not shown that
this is the only effect of $L_{1,2}$ for the other observables
that enter the global fits.
The calculation to ${\cal O}(1/\Lambda^4)$
can be made finite with the following renormalization of $\beta_1$:
\beq
\beta_1^r(\mu) = \beta_1
+{3 \over 4}{\alpha^2 (1 + c_Z^2)\over s_Z^2 c_Z^4}
\biggl(L_1 + {5 \over 2} L_2\biggr){v^2 \over \Lambda^2}
\biggl({1 \over \epsilon}+\log{\mu^2 \over M_Z^2}\biggr).
\label{shone}
\eeq

We obtain for the $Z$ partial widths:
\beq
{\delta\Gamma_f^{L_{1,2}} \over \Gamma_0(Z\ra f \overline{f})}=
-{3 \over 2}{\alpha^2 (1 + c_Z^2)\over s_Z^2 c_Z^4}
\biggl(L_1 + {5 \over 2} L_2\biggr){v^2 \over \Lambda^2}\log\biggl(
{\mu^2 \over M_Z^2}\biggr)
\biggl(1 +{2 r_f (l_f + r_f) \over l_f^2 + r_f^2}{c_Z^2 \over s_Z^2 - c_Z^2}
\biggr).
\label{resonetwo}
\eeq
Using $\Lambda=2$~TeV, and $\mu=1$~TeV we find $90\%$ confidence level bounds:
\beqn
\Gamma_e &\ra & -50 \leq L_1 + {5 \over 2}L_2  \leq 26 \nonumber \\
\Gamma_\nu &\ra & -28 \leq L_1 + {5 \over 2}L_2  \leq 59 \nonumber \\
R_h &\ra & -190 \leq L_1 + {5 \over 2}L_2 \leq 130 \nonumber \\
\Gamma_Z &\ra & -36 \leq L_1 + {5 \over 2}L_2 \leq 27
\quad .
\label{rhoana}
\eeqn
Combined, they yield the result:
\beq
-28 \leq L_1 + {5 \over 2} L_2 \leq 26
\label{combineonetwo}
\eeq
shown in Figure~1.

As mentioned before, the effect of $L_{1,2}$ in
other observables is very different from that of $\beta_1$.
It is
only for the $Z$ partial widths that we can make the ${\cal O}
(1/\Lambda^4)$ calculation finite with Eq.~\ref{shone}.

\subsection{Bounds on $L_{9L,9R}$ at order $1/\Lambda^4$.}

The calculation
of the effects of the $L_{9L}$ and $L_{9R}$ operators can
be carried out
simultaneously with the one-loop effects of the lowest
order effective Lagrangian, Eq.~\ref{lagt}, because
the form of the three and four gauge
boson vertices induced by these two couplings is the same as that
arising from Eq.~\ref{lagt}.
Performing the calculation in this way, we obtain a result
that contains terms of order $1/\Lambda^2$ (those independent of $L_{9L,9R}$),
terms of order $1/\Lambda^4$ proportional to $L_{9L,9R}$ and terms
of order $1/\Lambda^4$ proportional to $L_{10}$ and $\beta_1$.

\begin{figure}[htb]
\centerline{\hfil\epsffile{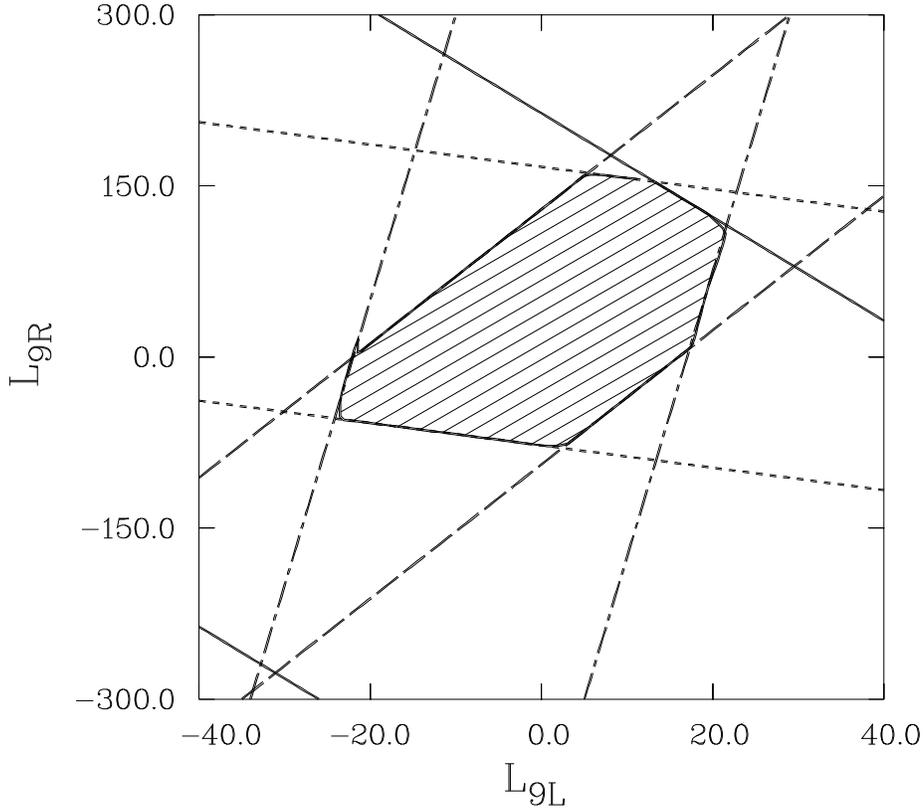}\hfil}
\caption[]{$90\%$ confidence level bounds on $L_{9L,9R}$ from the
$Z\rightarrow f {\overline f}$ partial widths, (Eq. 22).  The allowed
region is shaded.  The solid, dotted, dashed, and dot-dashed lines
are the bounds from $\Gamma_e$, $\Gamma_\nu$, $R_h$, and $\Gamma_Z$,
respectively.}
\end{figure}

It is amusing to note that the divergences generated by the operators
$L_{9L,9R}$ in the one-loop (order $1/\Lambda^4$) calculation of the
$Z \ra \overline{f} f$ widths can all be removed by the following
renormalization of the couplings in Eq.~\ref{oblique} (in the $M_t=0$ limit):
\beqn
\beta_1^r(\mu) &=& \beta_1 -
{\alpha \over \pi} {e^2\over 96 s_Z^4 c_Z^4}
{v^2 \over \Lambda^2}
\biggl[c_Z^2(1-20s_Z^2)L_{9L}+s_z^2(10-29c_Z^2)L_{9R}\biggr]
\biggl({1 \over \epsilon} + \log{\mu^2 \over M_Z^2}\biggr)
\nonumber \\
L^r_{10}(\mu)&=& L_{10} -
{\alpha \over \pi}{1\over 96 s_Z^2 c_Z^2}
\biggl[(1-24c_Z^2)L_{9L}+(32c_Z^2-1)L_{9R}\biggr]
\biggl({1 \over \epsilon} + \log{\mu^2 \over M_Z^2}\biggr)
\label{amusing}
\eeqn
This proves our assertion that our calculation to order ${\cal O}
(1/\Lambda^4)$ can be made finite by suitable renormalizations
of the parameters in Eq.~\ref{oblique}. However, we do not expect
this result to be true in general. That is, we expect that a
calculation of the one-loop contributions of the operators in
Eq.~\ref{lfour} to other observables will require counterterms
of order $1/\Lambda^4$. Thus, Eq.~\ref{amusing} does {\it not}
mean that we can place bounds on $L_{9L,9R}$ from global fits
to the parameters $S$ and $T$. Without performing a complete
analysis of the effective Lagrangian at order $1/\Lambda^4$
it is not possible to identify the renormalized parameters of
Eq.~\ref{amusing} with the ones corresponding to $S$ and $T$
that are used for global fits.

Keeping only terms
linear in $L_{9L,9R}$ we find,
\beqn
{\delta\Gamma^{L_9} \over \Gamma_0}&=&{\alpha^2 \over 24}
{1\over c_Z^4 s_Z^4}{v^2 \over \Lambda^2}
\log\biggl({\mu^2 \over M_Z^2}\biggr) \nonumber \\
&&\cdot \biggl\{ \biggl[L_{9L}(1-24c_Z^2)+L_{9R}(-1+32c_Z^2)\biggr]
{2r_f(l_f+r_f)\over l_f^2 +r_f^2}{c_Z^2 \over s_Z^2-c_Z^2}
\nonumber \\
&&+2\biggl[L_{9L}c_Z^2(1-20s_Z^2)+L_{9R}s_Z^2(10-29c_Z^2)\biggr]
\biggl(1+{2r_f(l_f+r_f)\over l_f^2 +r_f^2}
{c_Z^2 \over s_Z^2-c_Z^2}\biggr)\biggr\}
\nonumber \\
&&+{\alpha^2 \over 12}
\log\biggl({\mu^2 \over M_Z^2}\biggr)
{1+2c_Z^2 \over c_Z^4 s_Z^4 (l_b^2 +r_b^2)}
{v^2 \over \Lambda^2}
{M_t^2 \over M_Z^2}\biggl[L_{9R}s_Z^2-7L_{9L}c_Z^2\biggr]\delta_{fb} .
\label{finalres}
\eeqn
The last term in Eq.~\ref{finalres} corresponds to the non-universal
corrections proportional to $M_t^2$ that are relevant only
for the decay $Z \ra \overline{b} b$.

Using as before
$\Lambda=2$~TeV and  $\mu=1$~TeV we find $90\%$ confidence level bounds:
\beqn
\Gamma_e &\ra & -92 \leq L_{9L}+0.22L_{9R} \leq 47 \nonumber \\
\Gamma_\nu &\ra & -79 \leq L_{9L}+1.02L_{9R} \leq 170 \nonumber \\
R_h &\ra & -22 \leq L_{9L}-0.17L_{9R} \leq 16 \nonumber \\
\Gamma_Z &\ra & -22 \leq L_{9L}-0.04L_{9R} \leq 17
\label{rhonum}
\eeqn
We show these inequalities in Figure~2.

If we bound one coupling at a time
we can read from Figure~2 that:
\beqn
-22 \leq &L_{9L}& \leq 16 \nonumber \\
-77 \leq &L_{9R}& \leq 94 \quad .
\label{oneattime}
\eeqn
In a vector like model with $L_{9L}=L_{9R}$ we have
the $90\%$ confidence level bound:
\beq
-22<L_{9L}=L_{9R}< 18  .
\label{vectorbound}
\eeq

The limits from $\Gamma(Z\ra b {\overline b})$ are not competitive with those

\begin{figure}[htb]
\centerline{\hfil\epsffile{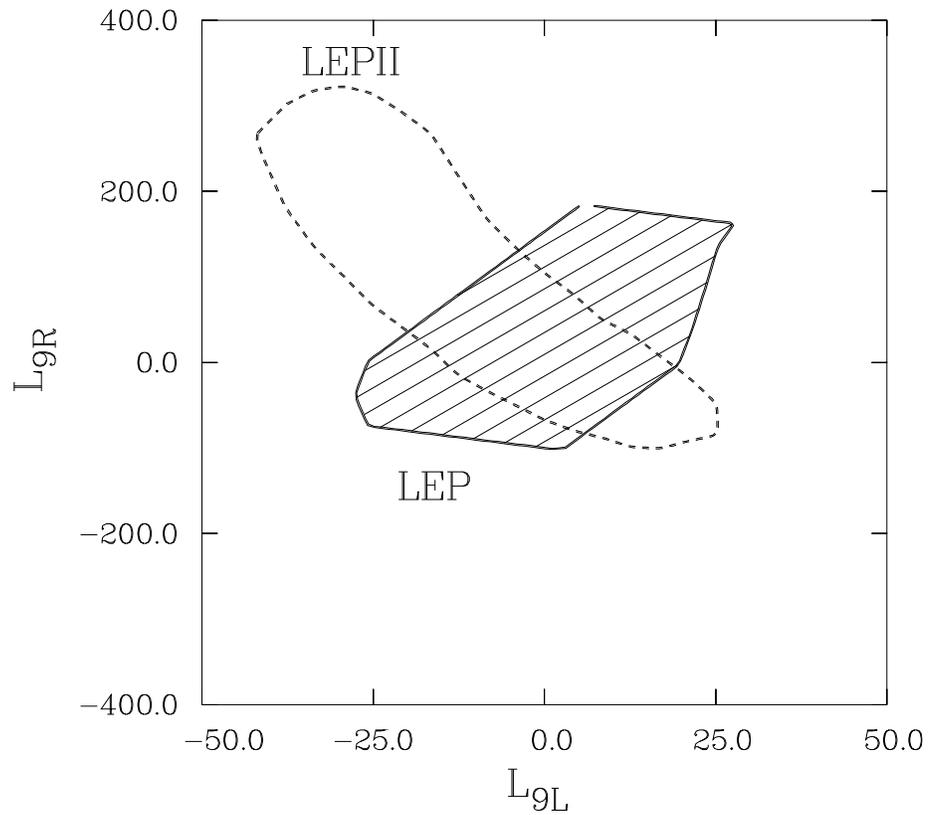}\hfil}
\caption[]{Comparison of the $95\%$ confidence level bounds from the
$Z$ partial widths (shaded region) with that obtainable at LEPII
with $\sqrt{s}=190~GeV$ and $\int {\cal L}=500~pb^{-1}$,
(dotted contour) \cite{boud}.}
\end{figure}
of Eq. \ref{rhonum} despite the enhancement of $M_t^2/M_Z^2$.

\subsection{Bounds on ${\hat \alpha}$ at order $1/\Lambda^4$}
The one loop contributions to $Z\ra f {\overline f}$ from the
parity violating operator ${\hat \alpha}$ cannot be made finite by
a renormalization of the operators of Eq. \ref{oblique}.
The divergence in this case will be absorbed by renormalization
of the couplings appearing at ${\cal O}(1/\Lambda^4)$.  The
leading non-analytic contribution can be extracted without
explicitely performing that renormalization.  It is,
\cite{gfive}
\beqn
{\delta \Gamma^5_f\over \Gamma_f^0}=
{3\alpha\over 2 \pi} g_5^Z\log\biggl({\mu\over M_W}\biggr)
\biggl[{2 l_f\over l_f^2+r_f^2}{c_Z^2\over s_Z^2}
+\biggl(1+{2 r_f(l_f+r_f)\over l_f^2+r_f^2}{c_Z^2
\over s_Z^2-c_Z^2}\biggr)\biggr] \quad ,
\label{gfans}
\eeqn
where
\beq
g_5^Z\equiv {4M_Z^2 {\hat \alpha}\over \Lambda^2}\quad .
\eeq

We find the results for $\Lambda=2~TeV$ and $\mu=1~TeV$,
\beqn
\Gamma_e &\ra & -.1 \leq g_5^Z \leq .05 \nonumber \\
\Gamma_\nu &\ra & -.08 \leq g_5^Z \leq .04 \nonumber \\
R_h &\ra & -.07 \leq g_5^Z \leq .1 \nonumber \\
\Gamma_Z &\ra & -.9 \leq g_5^Z \leq 1.2
\quad .
\label{g5num}
\eeqn

\section{Conclusion}

We can compare our results\footnote{
Our normalization of the $L_i$ is different from that
of Ref. [7].  We have translated their results into
our notation.}
 with bounds that future colliders are
expected to place on the anomalous couplings.

In Fig. 3, we compare
our $95\%$ confidence level bounds on $L_{9L}$ and $L_{9R}$ with those which
can be obtained at LEPII with $\sqrt{s}=190$~GeV and an integrated
luminosity of $500~pb^{-1}$. \cite{boud,bm2}
  We find that LEP and LEPII are
sensitive to slightly different regions of the $L_{9L}$ and $L_{9R}$
parameter space, with the bounds from the two machines being of
the same order of magnitude.
We again emphasize our caveat that
the  bounds from LEP rely on naturalness arguments and are no substitute
for measurements in future colliders.

The limits presented here on the four point couplings
$L_{1}$ and $L_{2}$ are the first available for these couplings.
They will be measured directly at the LHC.

Computing the leading contribution of each operator, and allowing
only one non-zero coefficient at a time, our $90~\%$ confidence level
bounds are:
\beqn
  -1.1 < &L_{10}^r(M_Z)_{new}& < 1.5 \nonumber \\
  -28 < &L_1& < 26  \nonumber \\
  -11 < &L_2& < 11  \nonumber \\
  -22 < &L_{9L}& < 16  \nonumber \\
  -77 < &L_{9R}& < 94  \nonumber \\
 -.07 < &g_5^Z& < .04 .
\label{rescon}
\eeqn
Two parameter bounds on ($L_1,L_2$) and ($L_{9L},L_{9R}$) are
given in the text.

\section{ Acknowledgements} The work of G. V. was supported in part
by a DOE OJI award. G.V. thanks the theory group at BNL for their
hospitality while part of this work was performed.
We are grateful to W. Bardeen, J.~F.~Donoghue, E. Laenen,
W. Marciano,  A. Sirlin, and A.~Sopczak for useful discussions.
We thank P.~Langacker for providing us with his latest numbers
and  F.~Boudjema for providing us with the data file for
the LEPII bounds in Figure~3.

\end{document}